\documentstyle[onecolumn]{mn}

\input{epsf}

\def\PP{{\cal P}}
\def\d{{\rm d}}
\def\ellq{l}

\begin{document}
\title[Solar sound-speed asphericity variations]{Solar-cycle variation 
of the sound-speed asphericity from GONG 
and MDI data 1995-2000}
\author[H.M. Antia et al.] {H.M. Antia$^1$, S. Basu$^2$,
F. Hill$^3$, R. Howe$^3$, R.W. Komm$^3$, J. Schou$^4$ \\
$^1$ Tata Institute of Fundamental Research, Homi Bhabha Road,
Mumbai 400005, India\\
$^2$ Astronomy Department, Yale University, P.O. Box 208101 New Haven,
CT 06520-8101, USA \\
$^3$ National Solar Observatory, 950 N. Cherry Ave., P.O. Box 26732, Tucson, Arizona,
85726-6732, USA \\
$^4$Hansen
Experimental Physics Laboratory, HEPL Annex A201, Stanford University,
Stanford, CA 94305-4085, USA}

\maketitle
\begin{abstract}
We study the variation of the frequency splitting coefficients
describing the solar asphericity in both GONG and MDI data, and
use these data to investigate temporal sound-speed variations 
as a function of both depth and latitude during the period from 1995-2000
and a little beyond.
The temporal variations in even splitting coefficients are found to
be correlated to the corresponding component of magnetic flux at the
solar surface.
We confirm that the sound-speed variations associated with the
surface magnetic field are superficial. 
Temporally averaged results show a significant excess in sound speed
around $r=0.92R_\odot$ and latitude of $60^\circ$.
\end{abstract}
\section{Introduction}

Helioseismology -- the study of acoustic oscillations in the Sun -- 
allows us to probe solar interior structure and rotation in 
two dimensions, depth and latitude, by taking 
advantage of the different spatial distribution of the various modes.
Recently, two projects -- the Global Oscillation Network Group 
(GONG) and the Solar Oscillations Investigation using the Michelson 
Doppler Imager (MDI) instrument aboard the {\it SOHO} spacecraft -- have provided
nearly five years of continuous helioseismic data, 
allowing the rise of the current solar cycle to be followed in 
unprecedented detail. 

Two-dimensional inversions for the rotation profile have been widely
studied. Inversions for the structural parameters (for example,
sound speed and density) are more commonly carried out in 
one dimension only.  However, it is possible to access 
two-dimensional information on these parameters also, using the
so-called `even $a$ coefficients'; results of early attempts can be 
found in Gough et al.~\shortcite{dogscience}.
In the light of recent work by Howe et~al.~\shortcite{ApJEpistle},
Howe, Komm and Hill \shortcite{HKH2000}, Toomre et~al. \shortcite{T2000}
and Antia and Basu \shortcite{AB00}
suggesting that the torsional-oscillation pattern seen at the surface
penetrates substantially into the convection zone, it is of
interest to see whether the structural changes also penetrate deeper
than has previously been thought.
In the work presented here we examine changes in helioseismic determinations
of the solar asphericity as the solar cycle progresses.

The modes are described by the radial order $n$, related to the number
of nodes in the radial direction, and the degree $l$ and azimuthal order $m$ 
which characterize the spherical harmonic defining the horizontal
structure of the mode. Rotation
and asphericity lift the degeneracy between the $2l+1$ modes of different
$m$ making up an ($n,l,m$) multiplet, resulting in frequency splitting.
The frequencies $\nu_{nlm}$ of the modes within a multiplet can be expressed 
as an expansion in orthogonal polynomials, for example
\begin{equation}
\nu_{nlm}
= \nu_{nl} + \sum_{j=1}^{j_{\rm max}} a_j (n,l) \, \PP_j^{(l)}(m). 
\label{eq:eq1}
\end{equation}

In early work (e.g., Duvall, Harvey \& Pomerantz, 1986) Legendre
polynomials were commonly used, whereas more recent work often uses the
Ritzwoller-Lavely formulation of the Clebsch-Gordan expansion,
where the basis functions are polynomials related
to the Clebsch-Gordan coefficients \cite{RL91}
$C_{j0lm}^{lm}$ by
\begin{equation}
\PP_j^{(l)}(m) = {l\sqrt{(2l-j)!(2l+j+1)!}\over
(2l)!\sqrt{2l+1}}C_{j0lm}^{lm}.
\label{eq:eq2}
\end{equation}

In either case, the coefficients $a_j$ are referred to as $a$-coefficients.
The odd-order coefficients, $a_1, a_3, \ldots$, are used to determine
the rate of rotation inside the Sun,
and reflect the advective, latitudinally symmetric part
of the perturbations caused by rotation. The even coefficients,
which are much smaller, are sensitive to second order contributions
from rotation, any possible magnetic field and any possible departure of the
solar structure from the spherically symmetric state.
Since the rotation rate in the solar interior is determined by the odd-order
coefficients, it can be used to calculate the second order contribution
of rotation to the even-order coefficients. Further, since the time variation in
the rotation rate is rather small, being of the order of 0.5\%, the
second order contribution due to rotation may be expected to be
essentially constant in time and hence any time-variation in the even-order
coefficients should be due to other sources, e.g., magnetic field
or asphericity. Unfortunately, it is not possible to distinguish
between these two possibilities using the even-order coefficients
(Zweibel and Gough 1995).


It was well established during the previous
solar cycle,
that the even coefficients 
show temporal variations related to solar activity measures.
The $a_2$ and $a_4$ coefficients of Duvall et al. \shortcite{DU86} 
were used by Gough and Thompson (1988)
to infer the possibility of a shallow magnetic perturbation in the
sound speed near the equator.
Kuhn (1988a) also 
pointed out the relationship between $a_2$ and $a_4$ measurements
from BBSO and earlier data and the latitudinal variation of internal 
sound speed, noting the possibility of
temporal variation as the known `hot bands' associated with magnetic activity
migrated during the solar cycle.
The temporal variation of the $a_2$ and $a_4$ coefficients and its
relation to changes in the latitudinal dependence of limb temperature 
measurements was further studied by Kuhn (1988b), Libbrecht (1989) and by  
Kuhn (1989), who predicted that the same relationship should extend to 
higher-order coefficients. 
The inference from the inversions of the BBSO even $a$ coefficient data 
carried out 
by Libbrecht and Woodard (1990) and Woodard and Libbrecht (1993)
was that most of the variation in the even coefficients was localized close to the surface and at the active latitudes, with a near-polar variation 
anticorrelated to the global activity level.
In the new cycle, we have data from  GONG and MDI
which allow us to study these trends in more detail.
Howe, Komm and Hill \shortcite{HKH99},
have found linear relations between 
the even-order $a$ coefficients and the Legendre decomposition components
of the surface magnetic flux up to $a_8$,
and showed that these relations could be extended backwards in time
to the BBSO data from the previous cycle. 
Dziembowski et al. (2000) found a good correlation between the 
even-$a$ coefficients derived from 12 72-day MDI data sets 
(1 May 1996 -- 31 May 1998) and the corresponding even components 
of the Ca II K data from Big Bear Solar Observatory (BBSO) up to 
$a_{10}$.

It has been pointed out by Kuhn \shortcite{K98} that the variations in mode frequency and
even $a$ coefficients are unlikely to be caused directly by the surface fields,
as the fields required would be substantially stronger than those
observed.
However, this does not rule out the idea that the presence of magnetic
fields affects the temperature. As pointed out by Dziembowski et al.~(2000),
the magnetic field might influence 
the thermal structure through the annihilation of the field 
($\beta$ effect) and indirectly through the perturbation of the 
convective transport ($\alpha$ effect) in addition to the direct 
mechanical effect of the Lorentz force.  
The good correlation between the
magnetic activity indices and the variation in mean frequencies and
even a-coefficients would tend to suggest that magnetic field
probably plays some role in these temporal variations.

In the present work we use more extensive data sets from both 
GONG and MDI, extending the GONG analysis up to $a_{14}$, and
present the results of 2-dimensional inversions for the sound speed.
We effectively assume that the even-order a-coefficients arise
from aspherical sound speed distribution, rather than being due to
possible magnetic field effects.

\section{Data}


We have analysed 53  overlapping 
108-day time series of GONG data
covering the period 1995 May 7 to 2000 October 6, centered
on dates 1 36-day `GONG month' apart. The data were 
analysed through the standard GONG pipeline \cite{DRAT96}, to find frequencies for each mode, and then 
$a$-coefficients were derived by fitting to the frequencies in 
each $(n,l)$ multiplet. A typical mode set from this process would 
consist of around 1400 multiplets with coefficients up to $a_{15}$,
for $l \leq 150$;
the intersection of all the sets contains about 400 multiplets mostly
with  $l \leq 100$. 

The MDI data consist of 23 non-overlapping time series
covering the period 1996 May 1 to 2001
April 4, although with some interruptions due to  problems
with the {\it SOHO} satellite. The data were analysed as 
described by Schou \shortcite {Sch99}.
These sets contain 
coefficients, up to $a_{36}$, for roughly 1800 multiplets with $l \le
300$.  The substantial temporal overlap between the two sets allows
useful cross-checking of results.

For the purpose of the inversions we have used only p-mode splittings
for modes with frequencies between 1.5 to 3.5 mHz. Further, only
coefficients up to $a_{14}$ are used as the higher order coefficients
do not appear to be significant. This gives us typically 8000 coefficients
for MDI data sets and 6000 coefficients for the GONG data sets.
Frequencies for f-modes ($n=0$) are also available for the MDI data,
and might provide additional information, but they have not been used in
the current work.

\section{Coefficient analysis}
\subsection{Method}

Howe et al.~\shortcite {HKH99} considered the temporal variation of the 
GONG mean even $a$ coefficients
up to $a_8$ and showed that they were strongly correlated with the
corresponding components of the Legendre decomposition of the 
surface magnetic field. In the present work we extend this 
analysis, with some modifications, to coefficients up to $a_{14}$
and to MDI as well as GONG data.

The latitudinal distribution of the surface magnetic flux can
be expressed as a sum of Legendre polynomials in $\cos\theta$,
\begin{equation}
B(t,\cos\theta) = \sum_k{B_k(t) P_k(\cos\theta)}.
\end {equation}
where $B_k(t)$ are time-varying coefficients. These coefficients can be
compared with the corresponding frequency splitting coefficients.
Global helioseismic measurements are sensitive only to the
latitudinally symmetric part of any departures from spherical symmetry ---
that is, to the even components of the expansion.

For the purposes of the present work, we consider two weighting schemes.
In the first, the splitting coefficients were
weighted by $E_{nl}$, where $E_{nl}$ is the mode inertia defined as in 
Howe et al. \shortcite{HKH99} 
and references therein, normalized to set the value for the radial 
mode at 3 mHz to unity.
In the second they were weighted by 
$l E_{nl}/Q_{lk}$, 
where $Q_{lk}$ is defined in eq.~\ref{eq:eq9} below,
before averaging over all common multiplets.
This removes the main $l$-dependent part of the coefficient variation, and
also reduces the frequency dependent part.
Thus, we define the quantities 
\begin{equation}
<a_{2k}>(t) = {{\sum_{n,l}{a_{2k}(n,l,t) E_{nl} \over {E_{nl}^2 \sigma_{a_{2k}}^2(n,l,t)}}}}
/{\sum_{n,l}{1\over{E_{nl}^2\sigma_{a_{2k}}^2(n,l,t)}}},
\label{eq:eq4a}
\end{equation}
and
\begin{equation}
<b_{2k}>(t) = {{\sum_{n,l}{a_{2k}(n,l,t) lE_{nl}/Q_{lk} \over l^2 {E_{nl}^2 \sigma_{a_{2k}}^2(n,l,t)}/Q_{lk}^2 }}}
/{\sum_{n,l}{Q_{lk}^2\over{l^2E_{nl}^2\sigma_{a_{2k}}^2(n,l,t)}}},
\label{eq:eq4b}
\end{equation}
where $t$ is time and $\sigma_{a_k}$ is the estimated uncertainty in the
coefficient $a_k$, and the sum is over the approximately 
400 $(n,l)$ multiplets that
are common to all the mode sets.

We then express the variation of the $<a_k>$
and $<b_k>$  as a function of the $B_k$,
for even $k$, as
\begin{equation}
<a_k>(t) = c_k + m_k B_k(t)
\end{equation}
and perform linear least-squares fits to obtain the gradient $m_k$ and
intercept $c_k$ for each $k$. The intercept will contain 
a contribution from the invariant part of the coefficients which
includes the second order effect of rotation,
as pointed out by Woodard \shortcite{Woodard89}, Gough and Thompson
\shortcite{GT90}, and 
Dziembowski and Goode \shortcite{DG92}.

\subsection{Results}

\begin{figure*}
\centerline{\epsfxsize=17cm\epsfbox{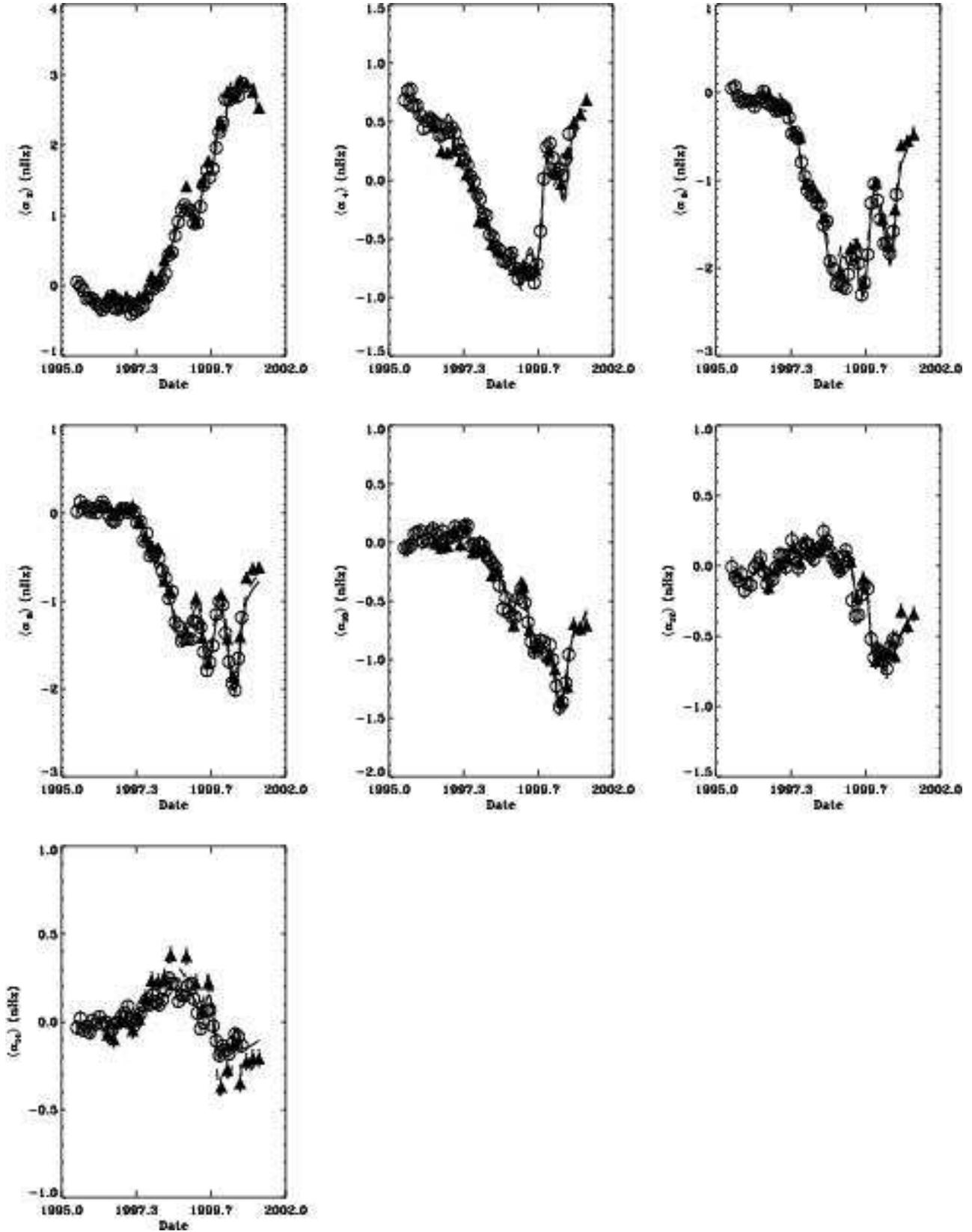}}
\caption{Temporal variation of the GONG (open circles) and MDI (filled triangles) 
mean even $a$ coefficients. The quantity plotted is $<a_k>$, 
as defined in Eq. \ref{eq:eq4a}.
The curves show the best-fit values
[$ m_k B_k + c_k$]   
for linear fits between the $<a_{2k}>$ coefficients and the 
corresponding $B_{2k}$ components of the Legendre decomposition of the
magnetic flux, for GONG (solid) and MDI (dashed).
}
\label{fig:coeff2}
\end{figure*}

In Figure \ref{fig:coeff2} we show the variations of the 
mean coefficients $<a_{2}>$ to $<a_{14}>$ as a function of time, for 
the sets of modes common to all datasets. 
The correlation between 
the coefficients and the Legendre components of the surface flux
persists up to $a_{14}$ in both sets of data (Table \ref{tab:tab1}). 
(The value of the correlation coefficient $R$ required for 
the 0.1 per cent significance level is 0.45 for GONG and 
0.68 for MDI.) There is a reasonably good agreement between GONG and MDI
data for $a_2,\ldots,a_{12}$, but for $a_{14}$ the agreement
is less good.  We suspect the difference may be
due to the different analysis procedures used for calculating the
splitting coefficients in the two data sets. 
In principle, $B_0$ may be correlated to the temporal variations in the mean
frequencies, but in practice, it is difficult to separate out the contributions to the mean
frequency due to the magnetic field from those due to uncertainties in the
spherically symmetric structure of the Sun. For this reason, we do not compare
the component $B_0$ in this work.

\begin{figure*}
\centerline{\epsfxsize=17cm\epsfbox{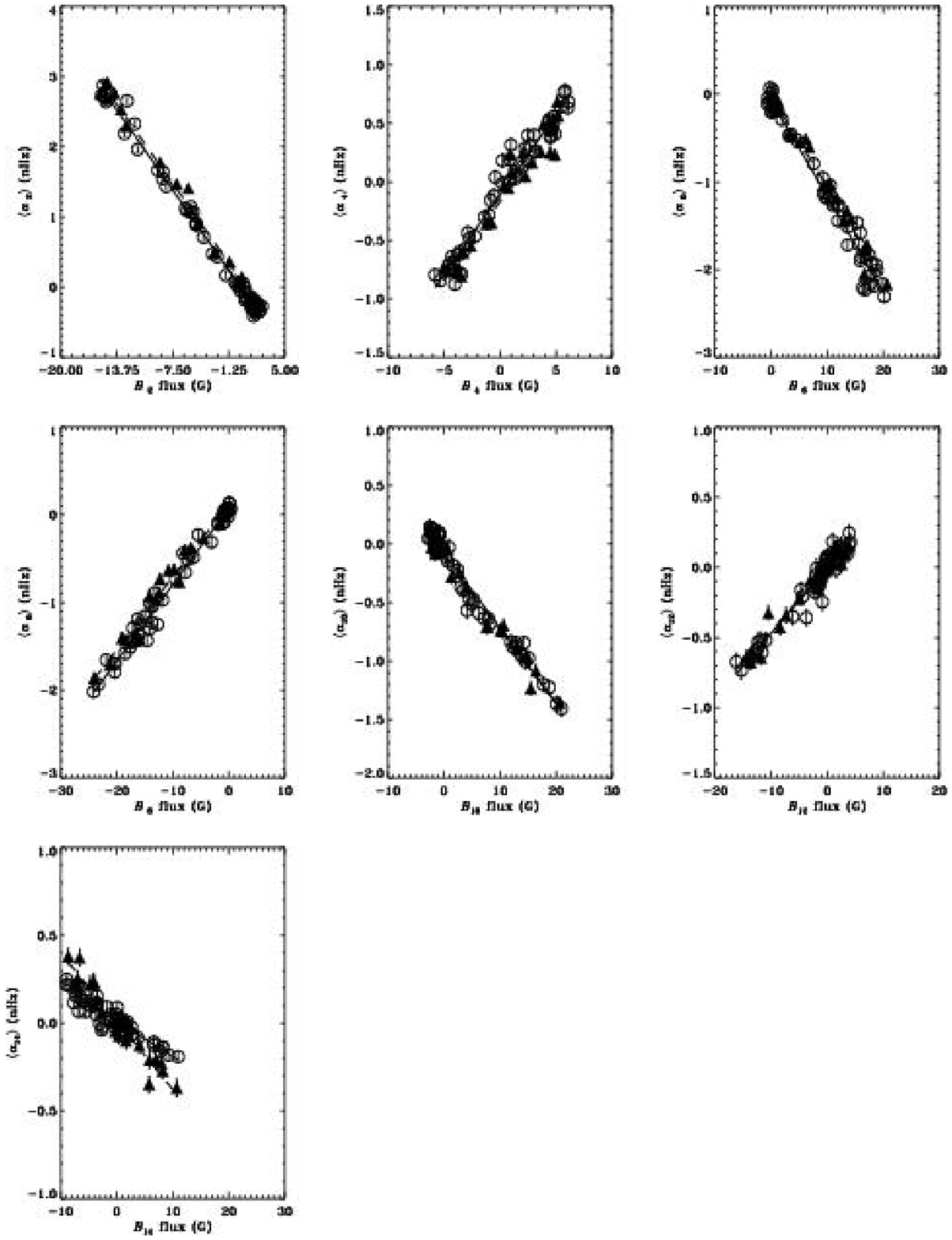}}
\caption{The relations between the GONG (open circles) and
MDI (filled triangles) $<a_{k}>$-coefficients 
and the corresponding $B_k$ components of the Legendre
decomposition of the magnetic flux from the Kitt Peak synoptic maps.
The lines show the best-fit results for linear fits to the 
data, for GONG (solid) and MDI (dashed).}
\label{fig:coeff1}
\end{figure*}

\begin{figure*}

\centerline{\epsfxsize=17cm\epsfbox{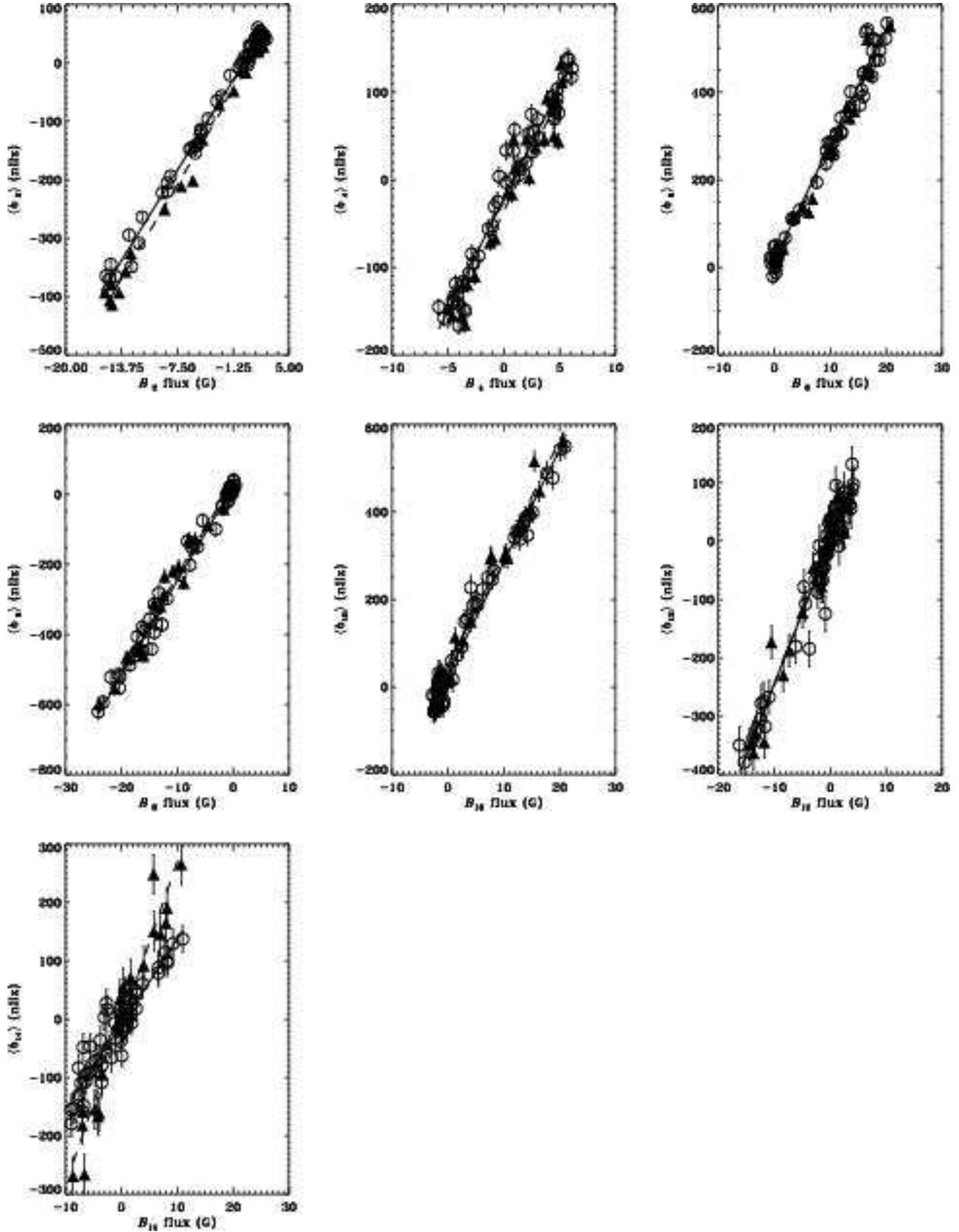}}
\caption{The relations between the GONG (open circles) and
MDI (filled triangles) $<b_{k}>$-coefficients 
and the corresponding $B_k$ components of the Legendre
decomposition of the magnetic flux from the Kitt Peak synoptic maps.
The lines show the best-fit results for linear fits to the 
data, for GONG (solid) and MDI (dashed).}
\label{fig:coeff1b}
\end{figure*}

In Figure \ref{fig:coeff1} we show the coefficients 
$<a_2>$ to $<a_{14}>$ as a function of the corresponding components
$B_{2}$ to $B_{14}$ of the surface magnetic flux, for GONG and MDI, 
while  the left portion of 
Table \ref{tab:tab1} gives the slopes $m_k$ and correlation coefficients
for the fits between even $a$-coefficients and their corresponding magnetic
flux decomposition components. 
While alternate even coefficients $<a_2>,<a_6>,<a_{10}>,\ldots$ appear to be
anticorrelated with $B_k$, the other coefficients $<a_4>,<a_8>,\ldots$ are
correlated with $B_k$. There also appears to be a reduction in magnitude of
$<a_k>$ with $k$ for $k>10$, while the $B_k$ do not show such a marked decrease.
However, the $<b_{2k}>$,  (Figure \ref{fig:coeff1b}, right side of 
Table \ref{tab:tab1}) are all positively 
correlated with 
$B_{2k}$ with scaling constants that appear to be independent of 
$k$, except for the highest order coefficients where the GONG data
show some reduction in sensitivity. 
In fact, most of the variation, including the sign changes, in the $<a_k>$ slopes up to $k=12$
can be explained by the corresponding variation in the angular integrals,
$Q_{lk}$.

\section{Asphericity inversions}
\subsection{Inversion techniques}

In order to study the variation of asphericity with depth and latitude we
apply an inversion technique to the even-order splitting coefficients.
We use the variational principle to analyse the
departures from a spherically symmetric solar model (e.g., Gough \shortcite{Gough93}),
in order to study aspherical perturbations to the sound speed and density
in the solar interior.
For simplicity, we only consider axisymmetric perturbations (with the
symmetry axis coinciding with the rotation axis) that are
symmetric about the equator. In this case, using the  variational
principle,  the difference in frequency between the Sun and a solar model for
a mode of a given order, degree and azimuthal order ($n$, $\ellq$, and $m$)
can be written as:
\begin{equation}
{\delta\nu_{nlm}\over\nu_{nlm}}=
\int_0^R\;\d r\;\int_0^{2\pi}\;\d \phi\;\int_0^\pi \sin\theta\;\d \theta\;
\left({\cal K}_{c^2,\rho}^{n\ellq}(r){\delta c^2\over c^2}(r,\theta)+
{\cal K}_{\rho,c^2}^{n\ellq}(r){\delta\rho\over\rho}(r,\theta)
\right)Y_\ellq^m(Y_\ellq^m)^*  
\label{eq:eqtest}
\end{equation}
where 
$r$ is radius, $\theta$ is colatitude,
${\delta\nu_{nlm}/\nu_{nlm}}$ is the relative frequency difference,
${\cal K}_{c^2,\rho}^{n\ellq}(r)$ and ${\cal K}_{\rho,c^2}^{n\ellq}(r)$
are the kernels
for spherically symmetric perturbations 
\cite{AB94},
and $Y_\ellq^m$ are spherical harmonics denoting the angular dependence of
the eigenfunctions for a spherically symmetric star. It is assumed that
the $Y_\ellq^m$'s are normalized such that
\begin{equation}
\int_0^{2\pi}\;\d\phi\;\int_0^\pi \sin\theta\;\d\theta\;Y_\ellq^m(Y_\ellq^m)^*
=1. 
\end{equation}
The perturbations ${\delta c^2/ c^2}$ and ${\delta\rho/\rho}$
can be expanded in terms of even order Legendre polynomials:
\begin{equation}
{\delta c^2\over c^2}(r,\theta)=\sum_k c_k(r)P_{2k}(\cos\theta), 
\end{equation}
\begin{equation}
{\delta \rho\over \rho}(r,\theta)=\sum_k \rho_k(r)P_{2k}(\cos\theta), 
\end{equation}
where $c_k(r)$ and $\rho_k(r)$ are shorthand notations 
for $({\delta c^2/ c^2})_k(r)$ and $({\delta\rho/\rho})_k(r)$ respectively.
The spherically symmetric component ($k=0$) gives frequency differences
that are independent of $m$ and thus only contribute to the mean frequency
of the $(n,\ellq)$ multiplet. Higher order terms give frequencies that
are functions of $m$ and thus contribute to the splitting coefficients.

The angular integrals in Eq. \ref{eq:eqtest} can be evaluated to give
\begin{equation}
\int_0^{2\pi}\;\d\phi\;\int_0^\pi \sin\theta\;\d\theta\;Y_\ellq^m(Y_\ellq^m)^*
P_{2k}(\cos\theta)={1\over\ellq}Q_{\ellq k}{\cal P}_{2k}^{(\ellq)}(m)
\label{eq:eq9}
\end{equation}
where $Q_{\ellq k}$ depends only on $\ellq,k$ and ${\cal P}_{2k}^{(\ellq)}(m)$
are the orthogonal polynomials defined by Eq. \ref{eq:eq2}.
The extra factor of $1/\ellq$ ensures that $Q_{\ellq k}$ approach a constant
value at large $\ellq$.
Thus with this choice of expansion (Eq. \ref{eq:eq2}) the inversion problem is decomposed
into independent inversions for each even splitting coefficient and
$c_k(r),\rho_k(r)$ can be computed by inverting the splitting
coefficient $a_{2k}$.
This would be similar to the  1.5d inversion to determine  the rotation rate
(Ritzwoller \& Lavely 1991),  called so  because
a two dimensional solution is obtained 
as a series of one dimensional inversions.

In practice, we also need to
account for the contribution to the frequency splittings that arises from
uncertainties in the treatment of
surface layers in the model. It is known that this contribution to
frequency splittings should be a slowly varying function of frequency alone
once it is corrected for differences in the mode
mass of the modes; no degree dependence is expected to a first approximation.
As in the case of inversions to determine the spherically symmetric 
structure of the Sun, the surface uncertainties are accounted for by
assuming that these can be represented by a frequency-dependent
function $F(\nu)$. However, in this case we determine a different
function $F_k(\nu)$ for each coefficient $a_{2k}$ and write the 
inversion problem as
\begin{equation}
{\ellq a_{2k}{(n,\ellq)}\over\nu_{n\ellq}}=
Q_{\ellq k}\int_0^R{\cal K}_{c^2,\rho}^{n\ellq}c_k(r)\;\d r+
Q_{\ellq k}\int_0^R{\cal K}_{\rho,c^2}^{n\ellq}\rho_k(r)\;\d r+
Q_{\ellq k}{F_k(\nu_{n\ellq})\over E_{n\ellq}}
\label{eq:eq10}
\end{equation}
Eq. \ref{eq:eq10} can be inverted using the usual inversion techniques to calculate
$c_k(r),\rho_k(r)$ and $F_k(\nu)$.
We use a Regularized Least Squares (RLS)
technique for this purpose and
expand the unknown functions $c_k(r),\rho_k(r),F_k(\nu)$
in terms  of cubic B-spline basis functions over knots,
which are approximately uniformly spaced in acoustic depth (or frequency).
First derivative smoothing is applied to constrain the error in the
solution to remain small.

As in the case of inverting  for rotation rate it is possible to
apply a 2d inversion technique for asphericity, where the asphericity is not
expanded in terms of Legendre polynomials, but instead one uses
\begin{equation}
{\ellq a_{2k}{(n,\ellq)}\over\nu_{n\ellq}}=
{Q_{\ellq k}(4k+1)\over2}\int_0^R\;dr\;\int_0^\pi\sin\theta\;d\theta
\left({\cal K}_{c^2,\rho}^{n\ellq}
{\delta c^2\over c^2}+{\cal K}_{\rho,c^2}^{n\ellq}{\delta \rho\over \rho}
\right)P_{2k}(\cos\theta)+
Q_{\ellq k}{F_k(\nu_{n\ellq})\over E_{n\ellq}}
\end{equation}
where ${\delta c^2/ c^2}$ and ${\delta \rho/ \rho}$ are
now functions of $(r,\theta)$ and can be expanded in terms of a
set of 2d basis functions for the RLS inversion. The knots are
chosen to be uniformly spaced in acoustic depth and $\cos(\theta)$
and we use the product of B-spline basis functions in $r$ and
$\cos(\theta)$. Thus, for example, we can write
\begin{equation}
{\delta c^2\over c^2}(r,\theta)=\sum_{i=1}^{n_r}\sum_{j=1}^{n_\theta}
b_{ij}\phi_i(r)\psi_j(\cos\theta),
\end{equation}
where $b_{ij}$ are the coefficients of expansion and $\phi_i(r)$
are the B-spline basis functions over $r$ and $\psi_j(\cos\theta)$
are those over $\cos\theta$; $n_r$ and $n_\theta$ are the number of
basis functions in $r$ and $\cos\theta$, respectively.
The unknown functions ${\delta c^2\over c^2}$,
${\delta \rho\over \rho}$ and $F_k(\nu)$ are expanded in terms
of basis functions and the coefficients of expansion are
determined by solving one inversion problem involving all splitting
coefficients. First derivative smoothing is applied in both $r$
and $\cos\theta$. Thus, the coefficients of expansion are
determined by minimizing
\begin{displaymath}
\sum_{n,\ellq,k}\left({\ellq\sigma_{n\ellq k}\over\nu_{n\ellq}}\right)^{-2}
\left[{la_{2k}{(n,\ellq)}\over \nu_{n\ellq}}-
\int_0^{R}\d r\int_{-1}^1\d\cos\theta\;
\left({\cal K}_{c^2,\rho}^{n\ellq k}(r,\theta){\delta c^2\over c^2}(r,\theta)
+{\cal K}_{\rho,c^2}^{n\ellq k}(r,\theta){\delta \rho\over \rho}(r,\theta)\right)
-{Q_{\ellq k}F_k(\nu)\over E_{n\ellq}}\right]^2
\end{displaymath}
\begin{displaymath}
\qquad +\lambda_r\int_0^{R}\d r\int_{-1}^1\d\cos\theta\;
r^{-1}\left(\left[{\partial\over\partial r} \left({\delta c^2\over c^2}\right)\right]^2+
\left[{\partial\over\partial r}\left({ \delta \rho\over\rho}\right)\right]^2\right)
\end{displaymath}
\begin{equation}
\qquad+\lambda_\theta\int_0^{R}\d r\int_{-1}^1\d\cos\theta\;
\sin^2\theta\left(\left[{\partial\over\partial\cos\theta}\left({\delta c^2\over c^2}\right)
\right]^2+
\left[{\partial\over\partial \cos\theta}\left({ \delta \rho\over\rho}\right)\right]^2
\right),
\end{equation}
where $\lambda_r$ and $\lambda_\theta$ are the two regularization
parameters controlling the smoothing and ${\cal K}_{c^2,\rho}^{n\ellq k}$
and ${\cal K}_{\rho,c^2}^{n\ellq k}$ are the 2d kernels,
$\sigma_{n l k}$ are the uncertainties in $a_{2k}{(n,\ellq)}$.
We have used 16 knots in $r$ and
10 knots in $\cos\theta$ to represent the asphericity.

\subsection{Inversion Results}

\begin{figure}
\centerline{\epsfysize=11cm\epsfbox{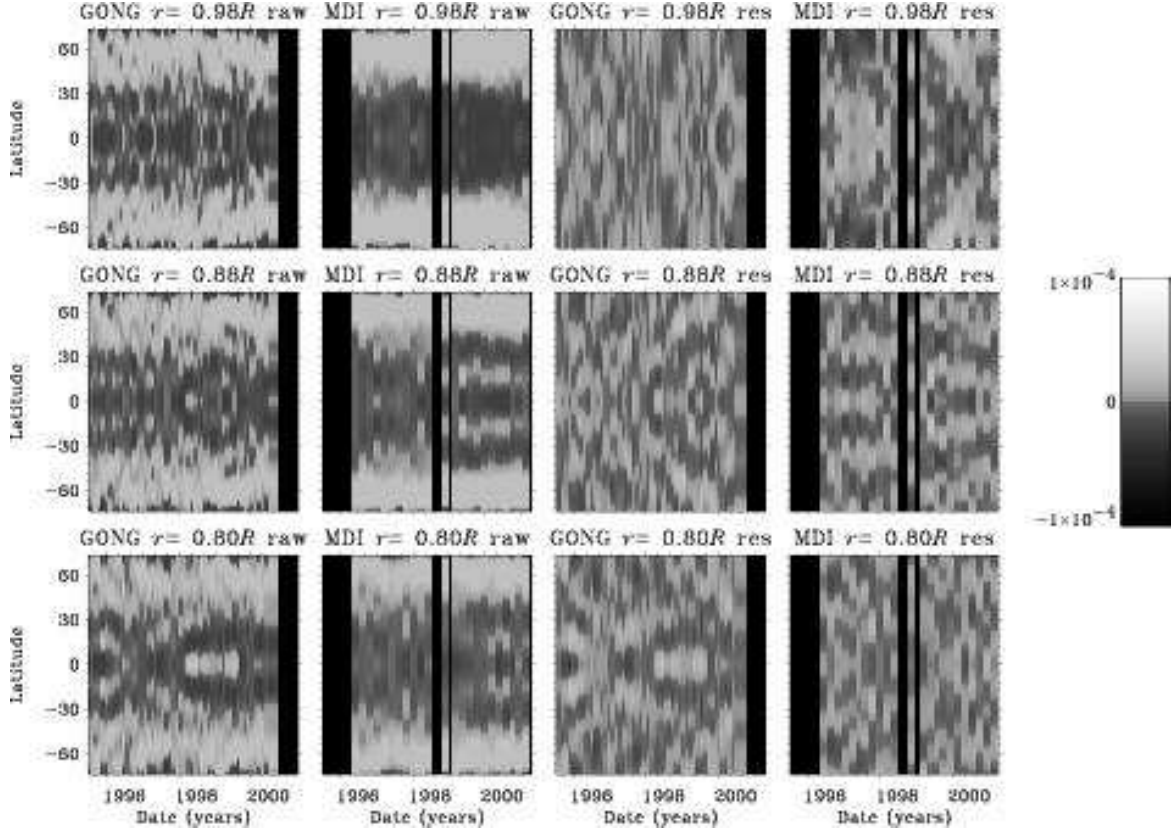}}
\caption{Grey scale maps showing the results of   
sound-speed inversions of GONG 
(first and third columns) and MDI (second and fourth columns) data 
at radii $0.98R$ (top), $0.88R$ (centre) and $0.80R$ (bottom). Blank 
spaces represent periods where there
are no data available. Columns 1 and 2 show the 1.5d inversion results
and columns 3 and 4 the 2d results.}
\label{fig:sixraw}
\end{figure}

In Figure \ref{fig:sixraw} we show the results of sound speed inversions
of GONG and MDI data at selected radii as a function of time and latitude.
There is no signature of significant temporal evolution at these radii. 
Both sets of data show a persistent sound-speed excess at around
60 degrees, apparently extending well down into the convection zone,
though this is more significant in the MDI than in the GONG data.

\begin{figure}
\centerline{\epsfysize=5cm\epsfbox{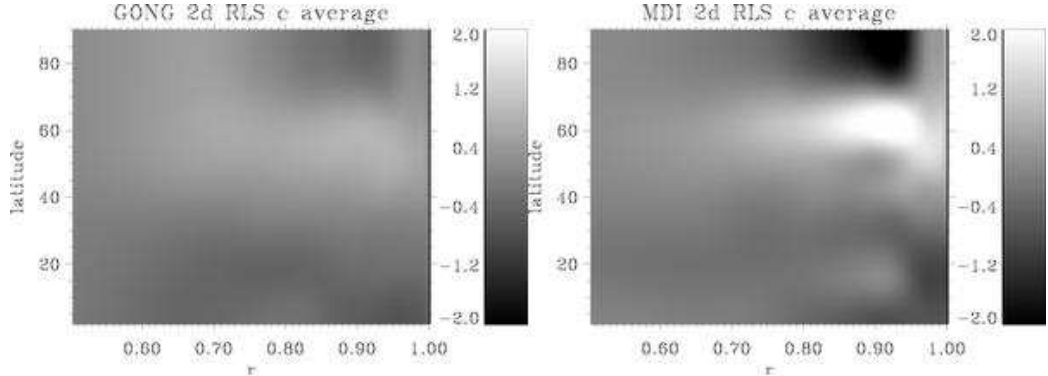}}

\caption{Grey scale maps showing the mean
results of sound-speed inversions of GONG 
(left) and MDI (right) data, multiplied by $10^4$,
as a function of latitude and radius.}
\label{fig:meanplot}
\end{figure}

The inversions show the deviation from a spherically symmetric model.
It turns out that the temporal mean of these deviations is non-zero
and has a dependence on depth and latitude, as illustrated in 
Figure \ref{fig:meanplot}.
However, it may be noted that the current data extend over only a
small fraction of a magnetic cycle and the temporal mean over this
limited period may not have much significance.
The second order contribution from rotation
would also contribute to the temporal mean.
There is some difference between the mean calculated from GONG and MDI
data sets, which could be due to systematic differences between the two
sets. In both cases there is a peak around depth of $0.08R_\odot$ and
the peak is more pronounced in MDI data.

\begin{figure}
\centerline{\epsfysize=11cm
\epsfbox{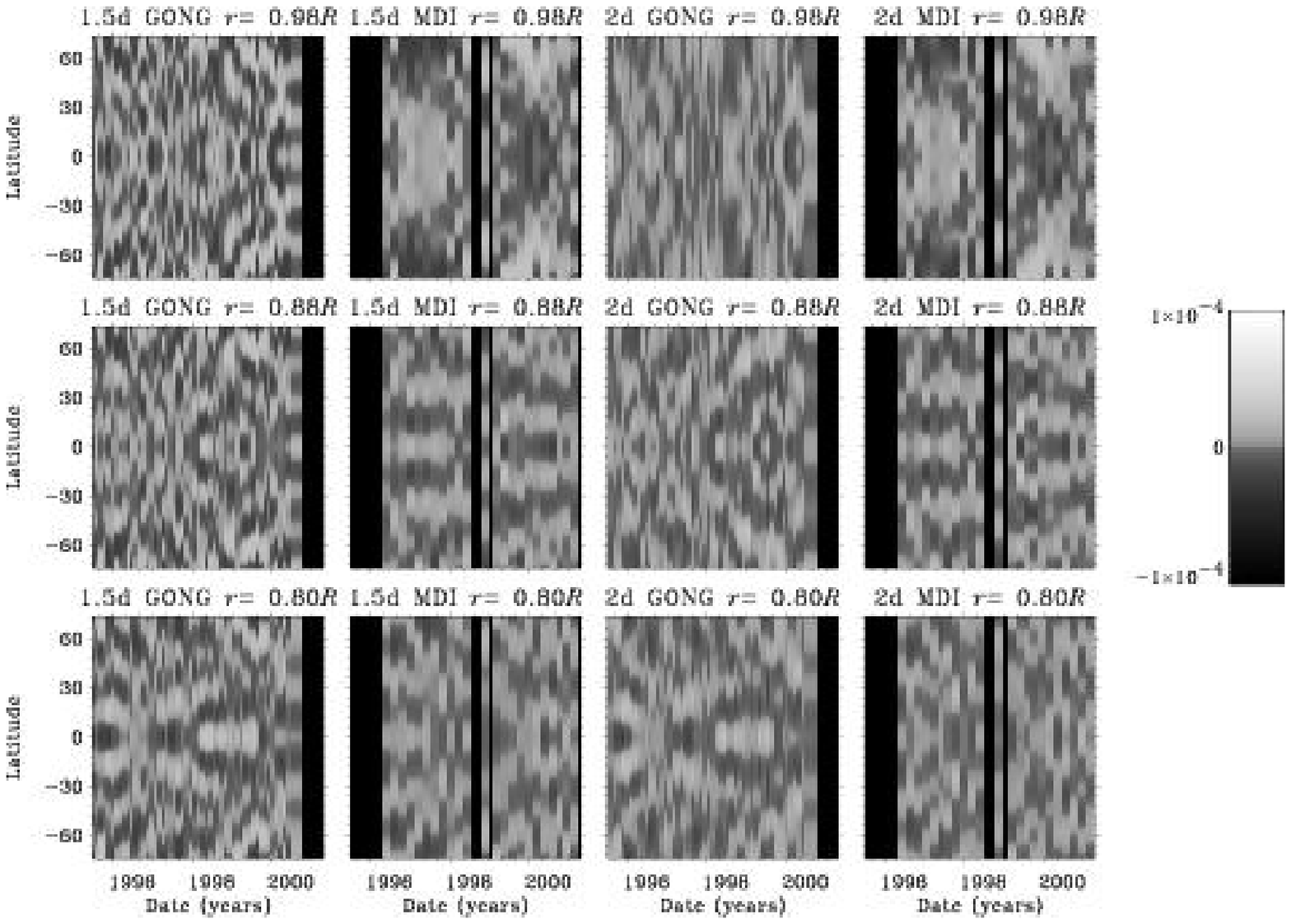}}
\caption{Grey scale maps showing the results of 
 sound-speed inversions of GONG 
(first and third columns) and MDI (second and fourth columns) data 
at radii $0.98R$ (top), $0.88R$ (centre) and $0.80R$ (bottom),
after subtraction of the temporal mean. Blank spaces represent periods where there
are no data available. Columns 1 and 2 show the 1.5d inversion results
and columns 3 and 4 the 2d results.}
\label{fig:sixres}
\end{figure}

\begin{figure}
\centerline{\epsfxsize=14cm\epsfbox{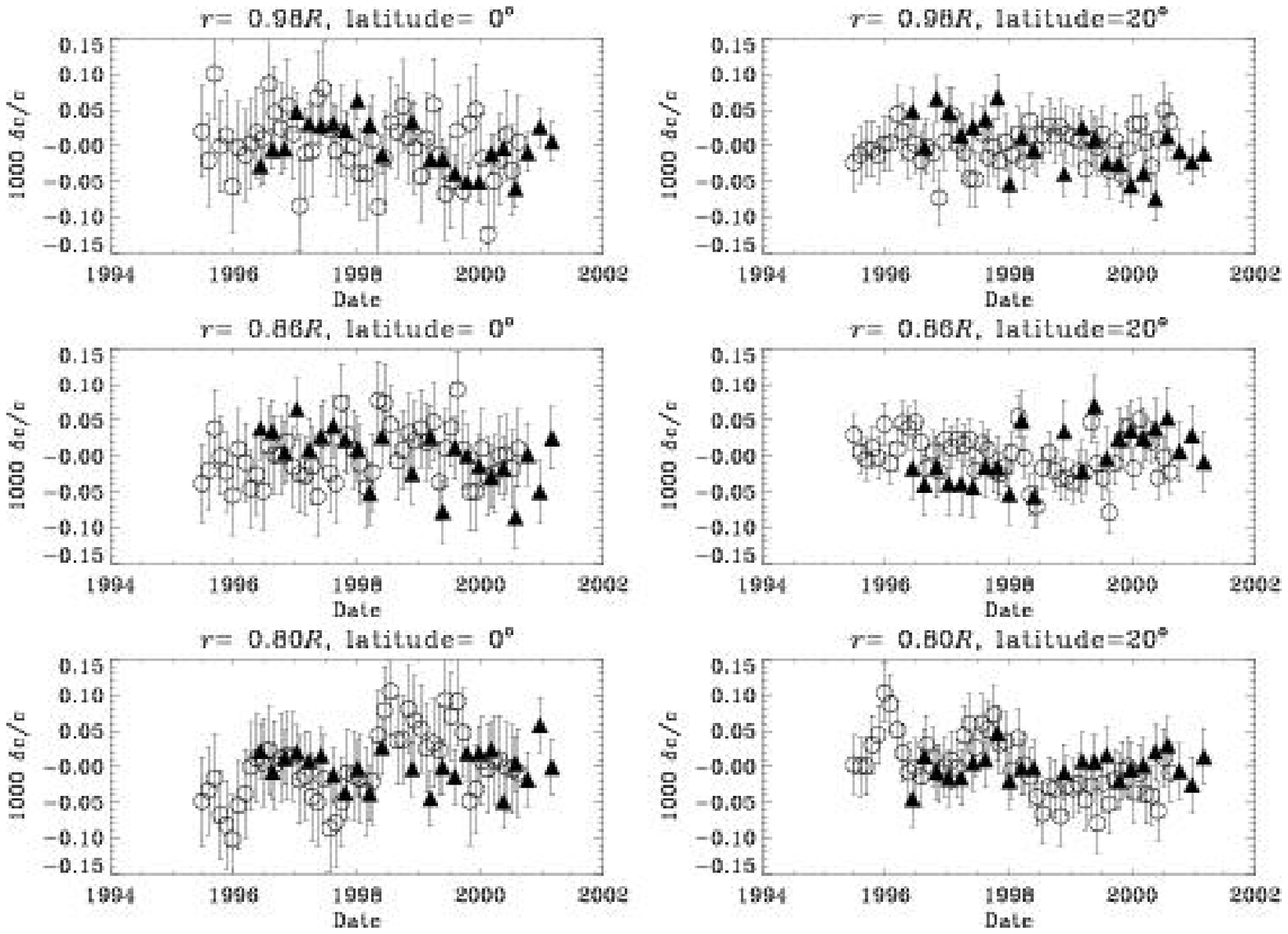}}
\caption{Variation of GONG (open) and MDI(filled) sound-speed
residuals at radii $0.98 R$ (top), $0.86 R$ (middle), and $0.8R$ (bottom)
for latitudes $0^{\circ}$ (left) and $20^{\circ}$ (right.)
}
\label{fig:bases}
\end{figure}

The sound-speed variation after subtraction of the mean profile is shown in 
Figure \ref{fig:sixres}. No systematic structure is evident.
Figure \ref{fig:bases} shows the variation of the sound-speed residuals
from the 2d inversions of
GONG and MDI data at selected $(r,\theta)$ points, illustrating the
agreement between the two 
experiments. The residuals are obtained by subtracting the temporal mean
from the results at each epoch.
It is clear from these figures that there is no significant
temporal variation in the asphericity at the depths we can resolve.

\begin{table}
\caption{Gradients $m$ and correlation coefficients $R$ for fits between 
the weighted mean coefficients $<a_{k}>$, $<b_{k}>$ and the corresponding 
Legendre components $B_k$ of the unsigned magnetic flux.}
\label{tab:tab1}
\begin{tabular}{@{}crrrrrrrr}
\ & GONG $\bar{a}_k$ & \ & MDI $\bar{a}_k$ & \ &GONG $\bar{b}_k$ & \ & MDI $\bar{b}_k$ & \ \\
$k$ & $m_k\ ({\rm nHz}) $ & $R$ & $m_k\ ({\rm nHz}) $ & $R$ & $m_k\ ({\rm nHz}) $ & $R$ & $m_k\ ({\rm nHz}) $ & $R$\\
       2 &$  -0.18\pm   0.00 $ & $   -1.00 $ & $   -0.18\pm   0.00 $ & $ 
  -1.00$ & $   24.81\pm   0.13 $ & $    1.00 $ & $   25.76\pm   0.16 $ & $ 
   1.00$ \\
       4 &$   0.14\pm   0.00 $ & $    0.98 $ & $    0.13\pm   0.00 $ & $ 
   0.96$ & $   25.57\pm   0.36 $ & $    0.98 $ & $   26.50\pm   0.57 $ & $ 
   0.96$ \\
       6 &$  -0.11\pm   0.00 $ & $   -0.99 $ & $   -0.10\pm   0.00 $ & $ 
  -0.99$ & $   26.50\pm   0.26 $ & $    0.99 $ & $   26.45\pm   0.43 $ & $ 
   0.99$ \\
       8 &$   0.09\pm   0.00 $ & $    0.99 $ & $    0.08\pm   0.00 $ & $ 
   0.99$ & $   27.06\pm   0.33 $ & $    0.99 $ & $   27.00\pm   0.51 $ & $ 
   0.99$ \\
      10 &$  -0.06\pm   0.00 $ & $   -0.99 $ & $   -0.06\pm   0.00 $ & $ 
  -0.99$ & $   25.70\pm   0.46 $ & $    0.99 $ & $   25.83\pm   0.64 $ & $ 
   0.99$ \\
      12 &$   0.05\pm   0.00 $ & $    0.97 $ & $    0.05\pm   0.00 $ & $ 
   0.98$ & $   24.40\pm   0.79 $ & $    0.97 $ & $   24.81\pm   0.98 $ & $ 
   0.98$ \\
      14 &$  -0.02\pm   0.00 $ & $   -0.93 $ & $   -0.04\pm   0.00 $ & $ 
  -0.97$ & $   14.56\pm   0.64 $ & $    0.93 $ & $   27.67\pm   1.30 $ & $ 
   0.97$ \\
\end{tabular}
\end{table}

\begin{figure}
\centerline{\epsfysize=5cm\epsfbox{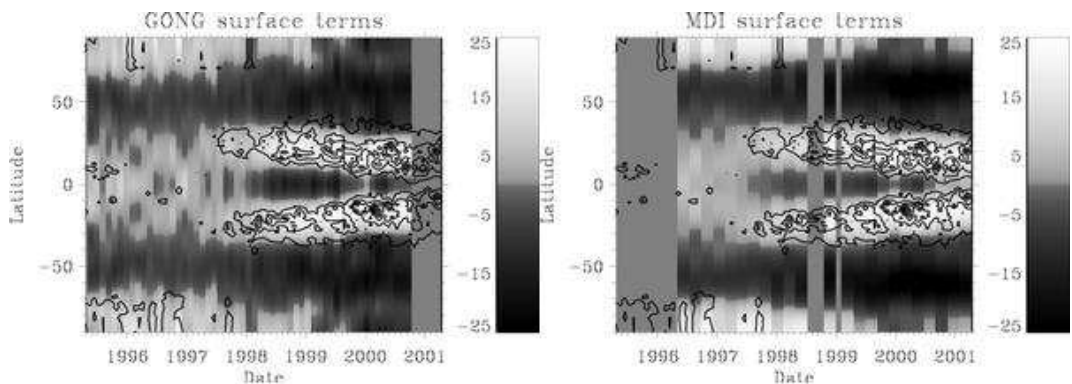}}
\caption{Grey scale map showing the reconstruction of the latitudinal dependence 
of the surface term from GONG (left) and MDI (right), multiplied by 100. Overlaid contours show the 
Kitt Peak unsigned magnetic flux with the $B_0$ term subtracted; contour spacing is 10G. 
}
\label{fig:surface}
\end{figure}

The strong variation in the even $a$ coefficients is evidently not reflected 
in the inversion results. Instead, it has been absorbed in the surface terms.
To illustrate this, we show in Figure \ref{fig:surface} the reconstruction 
of the surface term
for a frequency of 2.5 mHz, with overlaid contours of the magnetic flux.
As we would expect, since the dependence of the individual surface terms on 
frequency and on the magnetic field strength is largely independent of the order 
of the coefficient, this reconstruction matches the magnetic flux quite well.
This finding is consistent with the results of 
Woodard and Libbrecht (1993), who carried out latitudinal inversions of the
coefficients up to $a_{12}$. They found that the sound speed showed a 
peak at the active latitudes, matching both magnetic and temperature
behaviour, and also a decline at high latitudes with
increasing solar activity which was better matched by the temperature than 
by the magnetic data. The surface terms from our inversion show a similar
high-latitude behaviour, which does in fact correspond well to the
high-latitude magnetic data; we attribute this to an improvement in the
magnetic observations since 1992. Our inference that most of the solar-cycle 
variation in the sound speed is localized near the surface is 
also in agreement with the conclusions of Woodard and Libbrecht (1993).

\section{Resolution Issues}

In order to test our inversion method and to see how well
the inversions can resolve different features, we have
conducted tests with artificial data. Since asphericity inversions do
not include the spherically symmetric component corresponding to the mean
frequencies, $\delta c^2/c^2$ integrated over a surface with constant
$r$ must vanish. This arises because the non-vanishing part of the
integral can only
contribute to the mean frequencies which are not included. Furthermore, 
the even splitting coefficients are sensitive only to the north-south
symmetric component of the asphericity. Thus it is
necessary to choose artificial data which assume a profile that integrates
to zero in latitude and which are symmetric about the equator.
We have chosen a profile of the form
\begin{equation}
{\delta c^2\over c^2}=\cases{a_0\exp\left[-({r-r_0\over d})^2\right]
p_3(\cos^2\theta)& if $|r-r_0|<d$ \cr
0& otherwise \cr}
\end{equation}
where $r_0$ and $d$ are constants which respectively define the mean
depth and half-width of the aspherical perturbation and $p_3(\cos^2\theta)$
is a third degree polynomial in $\cos^2\theta$ which integrates to zero.
There is no special reason to choose a third degree polynomial, except that
with this choice one can get two bands of positive and negative asphericity
in each hemisphere. The amplitude $a_0$ of the signal is chosen to
approximately match the observed signal.

Using the artificial profile for $\delta c^2/c^2$ with the peak located at
different radii and with different widths, we construct a set of
artificial data. These data include only those modes which are present in
the observed data, and random errors with a Gaussian distribution with
standard deviation equal to the estimated uncertainties in observed data are added
to the calculated splittings.
The amplitude $a_0$ is chosen to be comparable to that obtained
from real solar data.
This process is followed for both GONG and MDI data sets.
These artificial data were then inverted the
way we would invert the real data and using the same choice for smoothing
as is used for real data. The inversion results can be compared
with the actual profile used in constructing the artificial data.

\begin{figure}
\centerline{\epsfxsize=14cm\epsfbox{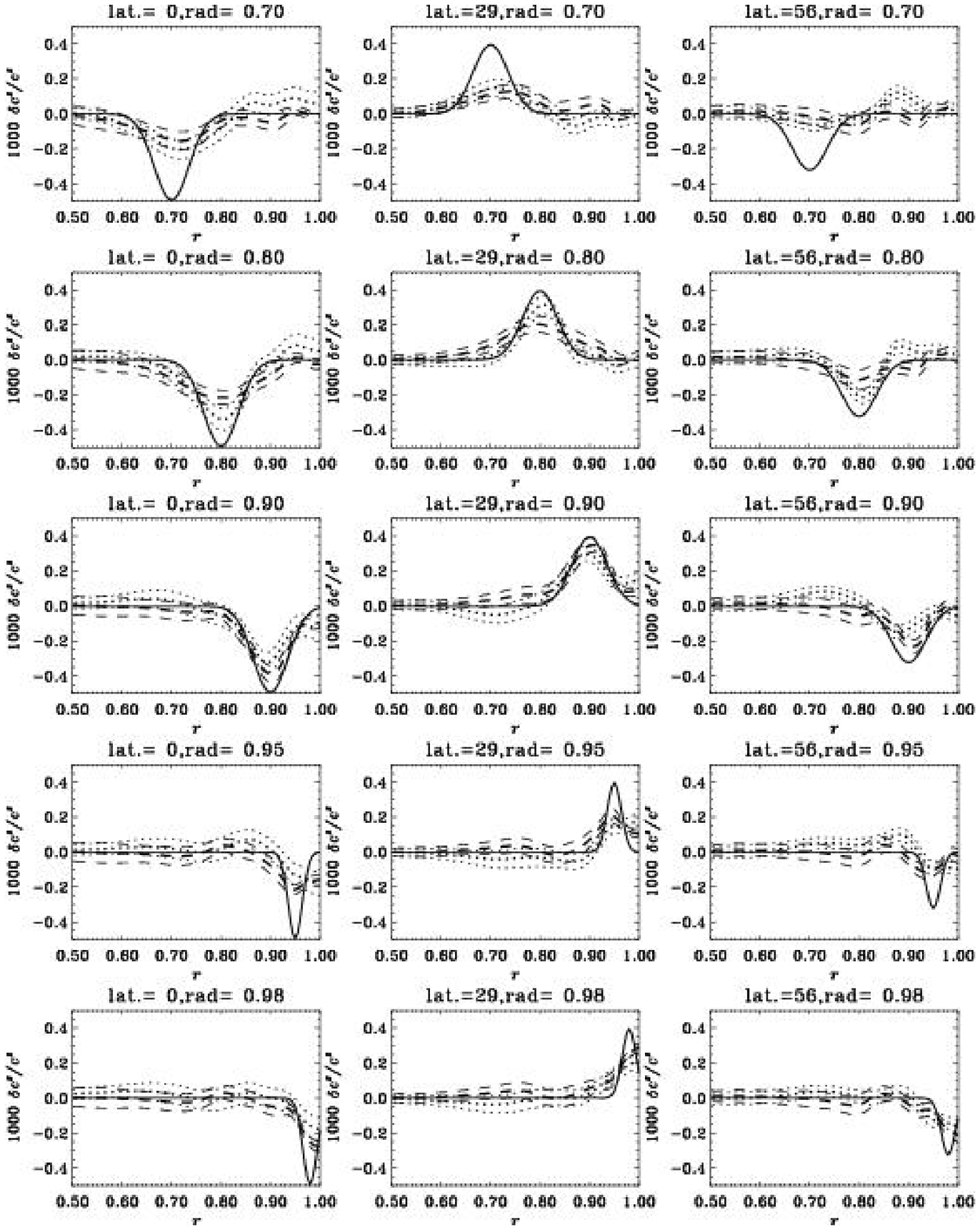}}
\caption{Artificial data inversion results for a variety of 
radial distances (from top to bottom, $r_0=0.7R$, $0.8R$, $0.9R$, $0.95R$, $0.98R$) and
latitudes (from left to right, latitude $0^\circ$, $30^\circ$ and
$56^\circ$), for GONG (dotted curves) and MDI (dashed curves.) The solid
curves show the `true' rate and the thinner dashed and dotted curves
represent 1-$\sigma$ error bounds on the inversion results.}
\label{fig:artslice}
\end{figure}

The inversion results show that we are able to invert for the
aspherical sound-speed difference in the outer layers of the Sun.
Fig.  \ref{fig:artslice}
shows the inversion results for artificial data with peak
located at different depths.
These results indicate that with the current data sets
we cannot invert for any 
feature located below a radius of 0.8 R$_\odot$.  At smaller radii,
e.g., $0.7R_\odot$, although we get some result at the equator,
the artificial profile  is not reproduced properly at high latitudes.
Similarly, the results become unreliable above
0.98 $R_\odot$, mainly because of the lack of high-degree modes.
The inversion results are
reliable between depths of $0.02R_\odot$ and about  $0.2R_\odot$.
As with all RLS inversions, there is
some structure in the results away from the peak, but the structure is
small and does not give rise to any major artifacts. The figures
shown give us confidence in the results we obtain by inverting real
data.

A similar exercise was also carried out for the inversions of density, 
but with an amplitude of artificial profile comparable to those in real data,
it was not found to be feasible to reproduce the profile to any reasonable
accuracy at all depths. Thus it appears that the aspherical component
of density cannot be reliably determined from current data sets.
Therefore, we do not show these results in the present work.

\section{Discussion}

We have studied the temporal variations in even a-coefficients from
GONG and MDI data during the period 1995--2001.
The mean values of the even $a$-coefficients (after scaling for mode mass and
angular integrals) over all modes are found to be well
correlated to the corresponding components of the observed magnetic
flux at the solar surface. The slope of best linear fit between the
mean splitting coefficient and the corresponding Legendre component of
the surface magnetic flux is found to be essentially independent of
the order of the coefficient. Thus the latitudinal variation in the
surface magnetic flux is correlated with that in asphericity as measured
by the even $a$-coefficients.
Dziembowski et al.\ (2000) obtained similar results analyzing 
12 72-day MDI data sets (1 May 1996 -- 31 May 1998).  They  
found a good correlation between the even-$a$ coefficients (up to 
$a_{10}$) and the corresponding even components of the Ca II K 
data from Big Bear Solar Observatory (BBSO).

In this work we have done two dimensional inversions for sound speed using
even a-coefficients from GONG and MDI data sets covering the
period from 1995 to early 2001, encompassing the
rising phase of the current solar cycle.
We find no significant temporal variation in the asphericity of the 
sound speed over this period. Thus it appears that the temporal variations
in even a-coefficients arise from changes taking place in surface layers.
A similar conclusion has been obtained for the spherically symmetric
component of sound speed and density \cite{BA00} as
the changes in mean frequency also appear to be associated with
surface effects.

For this work we have assumed that the even a-coefficients
arise from aspherical sound-speed distribution, but this is by no means
obvious as these coefficients could also arise from a magnetic field since both sets of 
kernels are very similar in most parts of the Sun and cannot 
be easily distinguished (see, for example, Figure~10 in 
Dziembowski et al.\ 2000).
In that case a magnetic field strength yielding $v_A^2/c^2\sim 10^{-4}$
would be required, where $v_A$ is the Alfven speed. Inside the convection
zone one might not expect an ordered magnetic field over large length scales
as turbulence might be expected to randomize the magnetic field. Such a
randomized magnetic field can also effectively change the wave propagation
speed, giving a signal similar to that from aspherical sound-speed
distribution. It may not be possible to distinguish between these
two effects from the even a-coefficients. Kuhn (1998) has argued that
the observed magnetic field at the solar surface is not sufficient to
explain the magnitude of the even $a$-coefficients. However, one can argue
that the magnetic field increases significantly as one goes deeper
in the near surface layers and this could explain the observed magnitudes
of the splittings.
 
The aspherical component of the density cannot be reliably determined
from the current data sets, but the magnitude is generally found to be smaller
than that for the sound speed. If the aspherical perturbations were of thermal
origin then one might have expected the density perturbations to be comparable
to the sound-speed perturbations.

Although the temporal variation in asphericity in the solar interior is
not significant, the mean asphericity is found to be significant and
shows a peak around $r=0.92R_\odot$ which is very pronounced in
the inversions of MDI data. This is deeper than the depth to which the
surface shear layer in solar rotation rate extends \cite{Sch98} and
about the same as the penetration depth recently 
(Howe et al.~2000; Antia and Basu 2000) established for the torsional oscillation pattern.
Antia, Chitre and Thompson (2000) attempted to detect a possible
signature for magnetic field using the even a-coefficients to find
that the coefficients $a_2$ and $a_4$ have a residual after
subtracting the contribution from rotation, which is peaked around
$r=0.95R_\odot$, which is consistent with our results.
Dziembowski et al.\ (2000) calculated inversions separately for 
temperature and magnetic field perturbations and detected a 
significant perturbation in the spherically symmetric part at a depth of 
25--100~Mm with a maximum at about 45~Mm, which agrees with 
the depth at which we find the aspherical perturbation in the
sound speed.
They interpret this perturbation as being due to 
either a magnetic perturbation of about (60 kG)$^2$ or a 
temperature perturbation of about 1.2~10$^{-4}$ which is 
of the same order as the result presented here.
Gilman (2000) pointed out that a thermal driving mechanism for the 
observed meridional flow would require a temperature excess at the
equator --- the opposite of what we observe. Thus this finding poses
yet another challenge for theoretical understanding of the dynamics
of the convection zone.

\section*{Acknowledgments}
This work utilizes data obtained by the Global Oscillation Network
Group (GONG) project, managed by the National Solar Observatory, which is
operated by AURA, Inc. under a cooperative agreement with the National
Science Foundation. The data were acquired by instruments operated by
the Big Bear Solar Observatory, High Altitude Observatory, Learmonth
Solar Observatory, Udaipur Solar Observatory, Instituto de Astrof\'{\i}sico
de Canarias, and Cerro Tololo Interamerican Observatory.
The Solar Oscillations Investigation (SOI) involving 
MDI is supported by NASA grant NAG 5-3077
to Stanford University.  {\it SOHO} is a mission of international cooperation
between ESA and NASA.  RWK, and RH in part, were supported by NASA
contract S-92698-F. 
NSO/Kitt Peak data used here are produced cooperatively by NSF/NOAO, NASA/GSFC, and NOAA/SEL.

\end{document}